\documentclass[a4paper]{article}
\usepackage{INTERSPEECH2021}  

\usepackage{graphicx} 

\usepackage{comment}
\usepackage{tikz}
\usepackage{color}
\usepackage{subfigure}      

\title{Comparison of remote experiments using crowdsourcing and laboratory experiments on speech intelligibility}

\name{ Ayako Yamamoto$^1$, Toshio Irino$^2$,   Kenichi Arai$^3$, Shoko Araki$^3$, Atsunori Ogawa$^3$, \\Keisuke Kinoshita$^3$, and Tomohiro Nakatani$^3$}
\address{
  $^{1,2}$ Faculty of Systems Engineering, Wakayama University,\\930 Sakaedani, Wakayama, 640-8510, Japan\\
   $^{3}$ 
   NTT Communication Science Laboratories,
  \\ 2-4 Hikaridai, Sekika-cho, Sorakugun,Kyoto, 619-0237, Japan
  }
\email{$^1$yamamoto.ayako@g.wakayama-u.jp, 
$^2$irino@wakayama-u.ac.jp, $^3$\{kenichi.arai.yw,shoko.araki.pu,atsunori.ogawa.gx,\\keisuke.kinoshita.mb,tomohiro.nakatani.nu\}@hco.ntt.co.jp}
\begin{document}
\maketitle
\begin{abstract}
\vspace{-5pt}
Many subjective experiments have been performed to develop objective speech intelligibility measures, but the novel coronavirus outbreak has made it very difficult to conduct experiments in a laboratory. One solution is to perform remote testing using crowdsourcing; however, because we cannot control the listening conditions, it is unclear whether the results are entirely reliable. 
In this study, we compared speech intelligibility scores obtained in remote and laboratory experiments. The results showed that the mean and standard deviation (SD) of the remote experiments' speech reception threshold (SRT) were higher than those of the laboratory experiments. However, the variance in the SRTs across the speech-enhancement conditions revealed similarities, implying that remote testing results may be as useful as laboratory experiments to develop an objective measure. We also show that the practice session scores correlate with the SRT values. This is a priori information before performing the main tests and would be useful for data screening to reduce the variability of the SRT distribution.
\end{abstract}


\vspace{-6pt}
\section{Introduction}
\vspace{-5pt}
Subjective speech intelligibility experiments provide fundamental information to develop objective intelligibility measures (e.g.,\cite{falk2015objective,yamamoto2020gedi}). They have been usually performed in a sound-proof room with well-controlled equipment in a laboratory. However, the novel coronavirus (COVID-19) outbreak has made it very difficult to conduct such formal experiments.
One solution is to perform remote testing with sound presentation and response collection using web pages. Although the participants can perform the experimental tasks at any location, it is almost impossible to control the acoustics and listening conditions, including their hearing levels. Hence, control is usually relinquished. Consequently, it is unclear whether the results are entirely reliable. This situation is a serious problem for any psychoacoustic experiments, and some good practices to overcome the issue were reported in \cite{ASARemoteTesting}.

However, from another point of view, remote testing using crowdsourcing is advantageous in collecting massive amounts of data from various participants when the control problem is not very serious. For example, it would be possible to analyze data categorized by listeners' characteristics if the volume of data is sufficiently large.
There have been many reports on using remote testing in speech quality assessments \cite{buchholz2011crowdsourcing, ribeiro2011crowdmos,naderi2015effect, jimenez2018influence, naderi2020open}.
In practice, remote testing has become popular in the quality assessment of text-to-speech synthesis algorithms. Particularly, it seems to be virtually a de facto standard in Interspeech competitions.
To improve reliability, methods for data screening were reported in order to reduce variability and eliminate false answers \cite{ribeiro2011crowdmos}.
Notably, there are relatively few studies about remote testing on speech intelligibility assessment \cite{cooke2011crowdsourcing, cooke2013crowdsourcing,paglialonga2020automated,padilla2021binaural}. It has not been common to perform remote testing with crowdsourcing because audio control problems and listeners' hearing levels remain crucial issues.

In this paper, we compared speech intelligibility results obtained from remote and laboratory experiments to verify whether remote testing is usable to develop objective intelligibility measures and to identify important factors toward improving the reliability of remote testing results.
	
\vspace{-6pt}
\section{Experiments: laboratory and remote}
\label{sec:ExpLabRemote}
\vspace{-5pt}
We performed web-based remote testing of speech intelligibility.
For precise comparison, the speech sounds for remote testing were basically the same as those used in laboratory experiments carried out to develop a new objective intelligibility model, GEDI  \cite{yamamoto2020gedi}. We briefly describe the speech materials that were common to the two experiments (see \cite{yamamoto2020gedi} for details) and explain the differences between them.

\vspace{-8pt}
\subsection{Speech materials}
\label{sec:SpeechMaterial}
\vspace{-5pt}
The speech sounds used for the subjective listening experiments were Japanese 4-mora words, spoken by a male speaker (label ID: mis), from a database of familiarity-controlled word lists, FW07 \cite{Kondo2007}. Note that one mora in Japanese roughly corresponds to a vowel or a consonant-vowel (CV) syllable and is written as a single hiragana character, except for some minor examples\cite{Mora}. The database comprises several word-familiarity ranks corresponding to the degree of lexical information. 
Speech sounds were obtained from the sound set with which the participants were the least familiar to prevent listeners from complementing their answers with guesses. The dataset contains 400 words per single familiarity, and the average duration of a 4-mora word is approximately 700\,ms. 

Babble noise was added to the clean speech to obtain noisy speech sounds, referred to as ``unprocessed.'' The SNR conditions ranged from $-6$ to $+6$\ dB in $3$-dB steps, and the duration was adjusted to the original speech sound.  
Two speech enhancement algorithms were applied to the unprocessed sounds. The first was a simple spectral-subtraction (SS) algorithm \cite{Berouti1979}. With an over-subtraction factor of 1.0, it is referred to as ``${\rm SS^{(1.0)}}$.''
The second one was a Wiener filter (WF) based algorithm that is commonly used in various systems because of its effectiveness with low computational costs. In particular, the WF based on a pre-trained speech model (PSM) \cite{Fujimoto2012} was used in \cite{yamamoto2020gedi}.
With Wiener gain parameter values of 0 and 0.2, the WF using the PSM is referred to as ``$\rm WF_{PSM}^{(0.0)}$'' and ``$\rm WF_{PSM}^{(0.2)}$,'' respectively.  
All noise addition and speech enhancement processes were performed at a sampling rate of 16\,kHz, and the final sounds delivered to the listeners were re-sampled to 48\,kHz.

\vspace{-8pt}
\subsection{Laboratory experiments}
\label{sec:ExpLab}
\vspace{-5pt}
\color{black}
In the laboratory experiments\cite{yamamoto2020gedi}, the sounds were presented diotically via a DA converter (OPPO, HA-1) over headphones (OPPO, PM-1) at a sampling frequency of 48\,kHz and a quantization level of 24 bits. Sound presentation was controlled using MATLAB in Mac OS X. The sound pressure level (SPL) of the stimulus sounds was 63\,dB in ${L_{\rm Aeq}}$. These laboratory experiments are referred to as having a moderate SPL. Listeners were seated in a sound-attenuated room with a background noise level of approximately 26\,dB in $L_{\rm Aeq}$.

Fourteen young NH listeners (eight males and six females, aged between 19 and 24 years) participated in the experiments. The participants had a hearing level of less than 20\,dB between 125 and 8,000\,Hz, and their native language was Japanese. 
They participated in the experiments only after providing informed consent. The participants were instructed to write down the words they heard using hiragana during a 4-second silent period until the presentation of the next word.  The total number of presented stimuli was 400 words, comprising a combination of four speech-enhancement conditions  \{``Unprocessed'', ``${\rm SS^{(1.0)}}$'', ``${\rm WF_{PSM}^{(0.0)}}$'', and ``${\rm WF_{PSM}^{(0.2)}}$'' \} and five SNR conditions with 20 words per condition.
The total duration of the listening test was about 1 hour. To keep the listeners' attention within a reasonable range, we restricted the maximum number of words to 400 in order to cover all SNR conditions and enhancement algorithms. 
Each subject listened to a different word set, which was assigned randomly to avoid bias caused by word difficulty. 

We also performed complementary low SPL laboratory experiments, 43\,dB in ${L_{\rm Aeq}}$ (i.e., a -20\,dB reduction from the above experiment), to estimate the effect of the SPL on speech intelligibility \cite{Iwaki2019effect}.
Another 14 NH listeners participated in the experiments. The speech materials and procedures were essentially the same, except the SNR conditions ranged from $0$ to $+12$\ dB in $3$-dB steps, since we assumed a decrease in intelligibility due to the low SPL. 

\vspace{-8pt}
\subsection{Remote experiments}
\label{sec:ExpRemote}
\vspace{-5pt}
The remote experiments were performed using web pages that had been newly developed for speech intelligibility tests\cite{WebSpIntel}. 
To reduce the experiment duration to within 1 hour, we divided the speech-enhancement conditions into two parts: \{``Unprocessed'', ``${\rm SS^{(1.0)}}$''\} and \{``${\rm WF_{PSM}^{(0.0)}}$'', ``${\rm WF_{PSM}^{(0.2)}}$''\}. Each set consisted of 200 non-overlapping words, i.e., 10 words x 20 sessions.  All participants listened to the same word set, since dynamic assignment was unavailable on the web pages. 

The participants were instructed to write down the words they heard using hiragana during a 4-second silent period until the presentation of the next word. The answers were filled in on the provided answer sheets (PDF), which had been printed in advance. After listening to ten words (i.e., one session), the participants were required to type the hand-written words into the answer columns on the web page. 

The experimental tasks were outsourced to a crowdsourcing service provided by Lancers Co. Ltd. in Japan \cite{Lancers}, where, it is claimed that 100,000 workers are registered, with their personal information, including skills. We recruited 30 participants per experiment without specifying any conditions regarding age, gender, hearing level, and educational background. The only requirements were to use a personal computer and wired headphones or wired earphones, to avoid Bluetooth devices and loudspeakers. As a result, there was a large variety of participants aged from their mid-20s to 60 years old.
The first experiment was finished within 2 days. The second experiment opened a few days after. Any worker can participate in the experimental task on a first-come-first-served basis. Since we wanted the same workers to participate in both experiments, we added some advertising phrases about the second experiment at the end of the first one. Sixteen workers participated in both, to a total of 44 participants.

Initially, on the web pages, the participants were required to read information about the experiments before giving informed consent by clicking the agreement button twice in order to be transferred to the experimental task web page. Google Chrome was specified as a usable browser because it plays wav files with 48-kHz and 16-bit properly in both Windows and Mac. The participants set their devices at an easily listenable level. 

To familiarize with the experimental tasks, the participants took a training session in which they performed a very easy task using the same procedure as in the test sessions. The speech sounds were drawn from words in the highest familiarity rank with an SNR above 0 dB.
After the analysis, it was found that this practice session may play an important role in data screening, as described in section \ref{sec:DataScreening}.

\begin{figure*}[t]
      \vspace{-10pt}
  \begin{center}
    \begin{tabular}{c}
      \begin{minipage}{0.65\columnwidth}
        \begin{center}
        \hspace{-12mm}
          \includegraphics[width=1.2\linewidth]{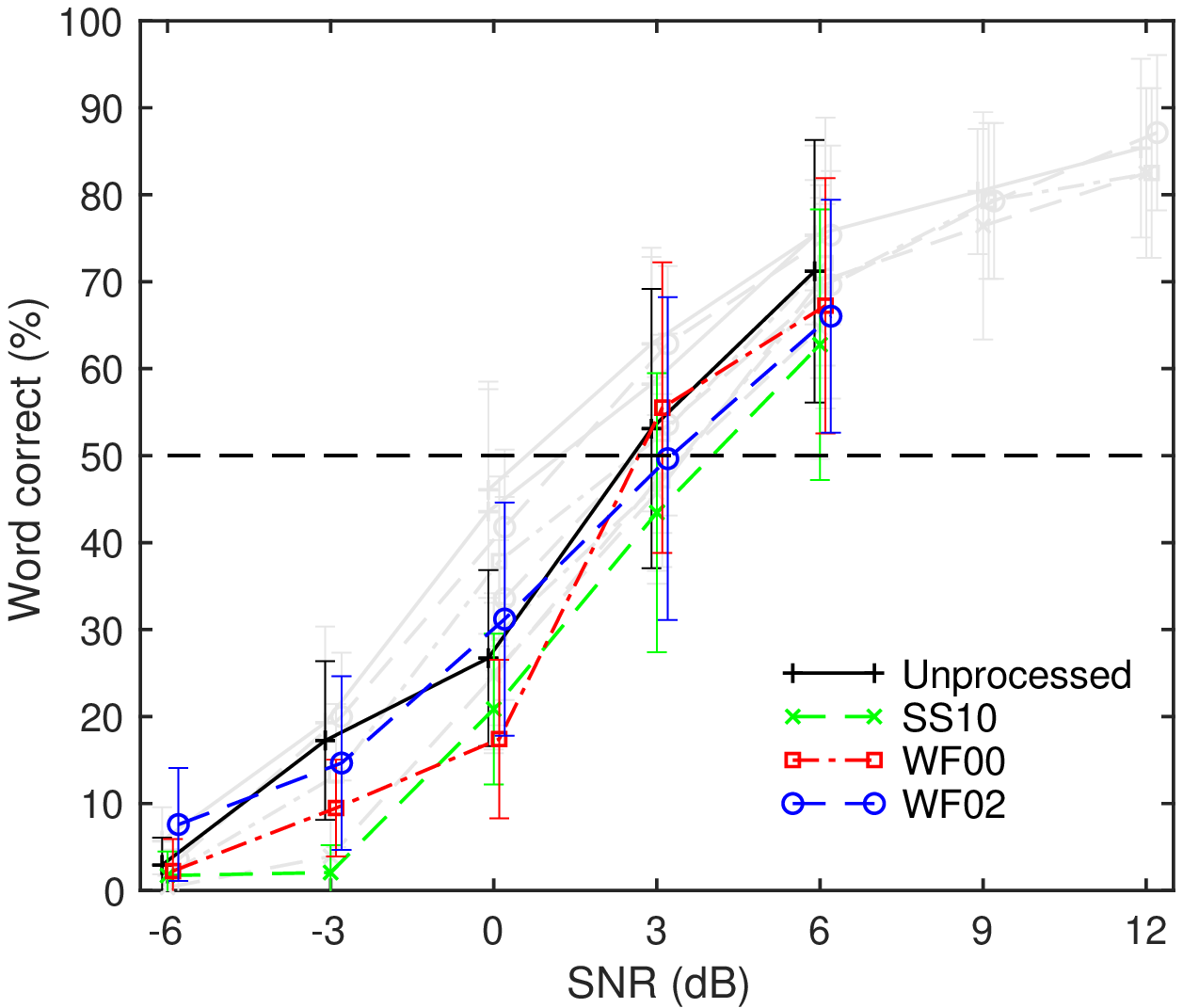}
          \caption{Crowdsourcing remote experiment. Mean and SD of word correct (\%). SS10: ``${\rm SS^{(1.0)}}$'',  WF00: ``${\rm WF_{PSM}^{(0.0)}}$'', WF02: ``${\rm WF_{PSM}^{(0.2)}}$''. }
          \label{fig:ExpRemote_WCcurve}
        \end{center}
      \end{minipage}
    \hspace{7mm}
      \begin{minipage}{0.65\columnwidth}
        \begin{center}        
        \hspace{-12mm}
          \includegraphics[width=1.2\linewidth]{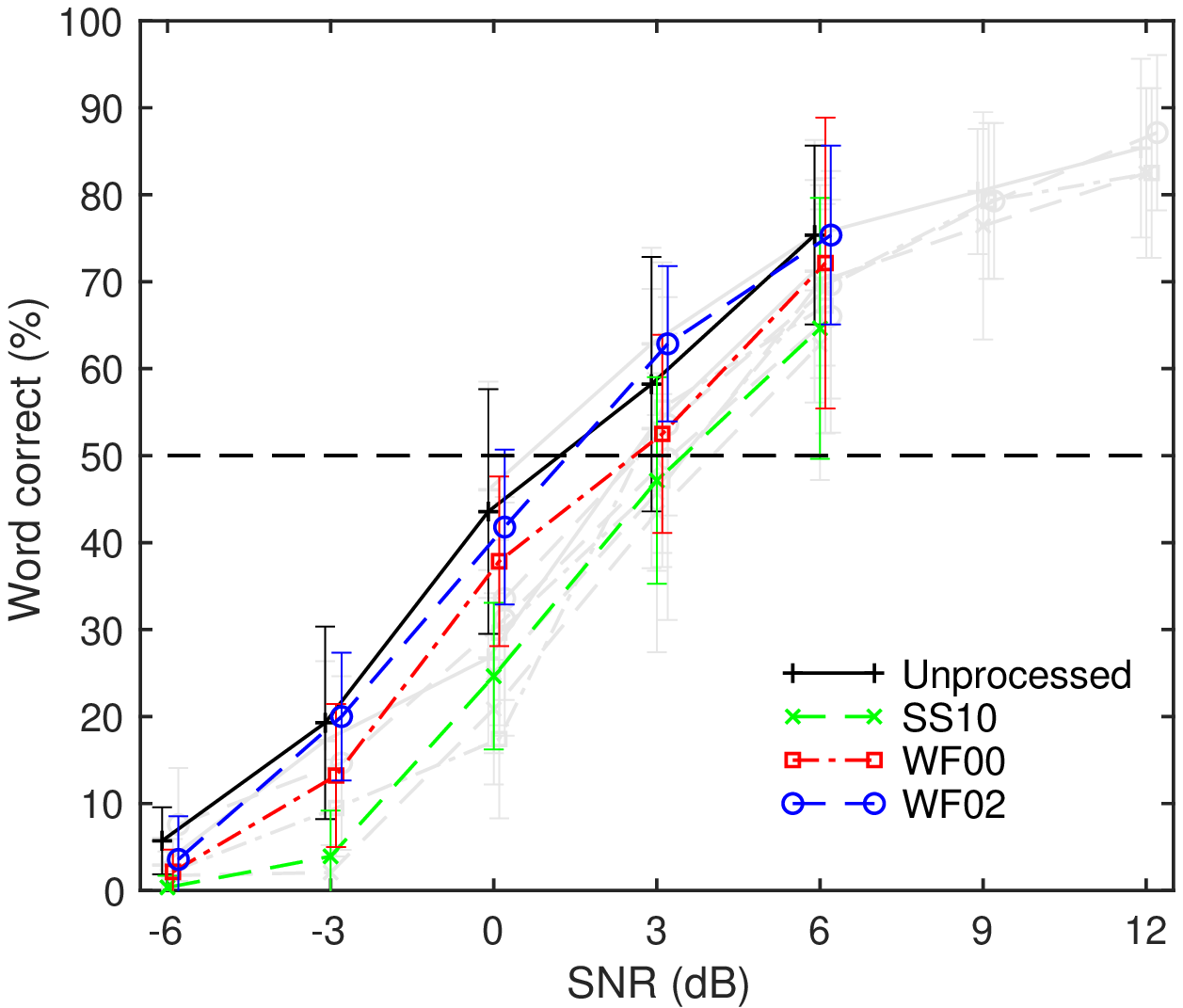}
          \caption{Laboratory experiment (moderate SPL, 63 dB $L_{\rm Aeq}$)\cite{yamamoto2020gedi}. Mean and SD of word correct  (\%)  \\}
            \label{fig:ExpLab_WCcurve}
        \end{center}
      \end{minipage}
    \hspace{7mm}
      \begin{minipage}{0.65\columnwidth}
        \begin{center}
                \hspace{-12mm}
          \includegraphics[width=1.2\linewidth]{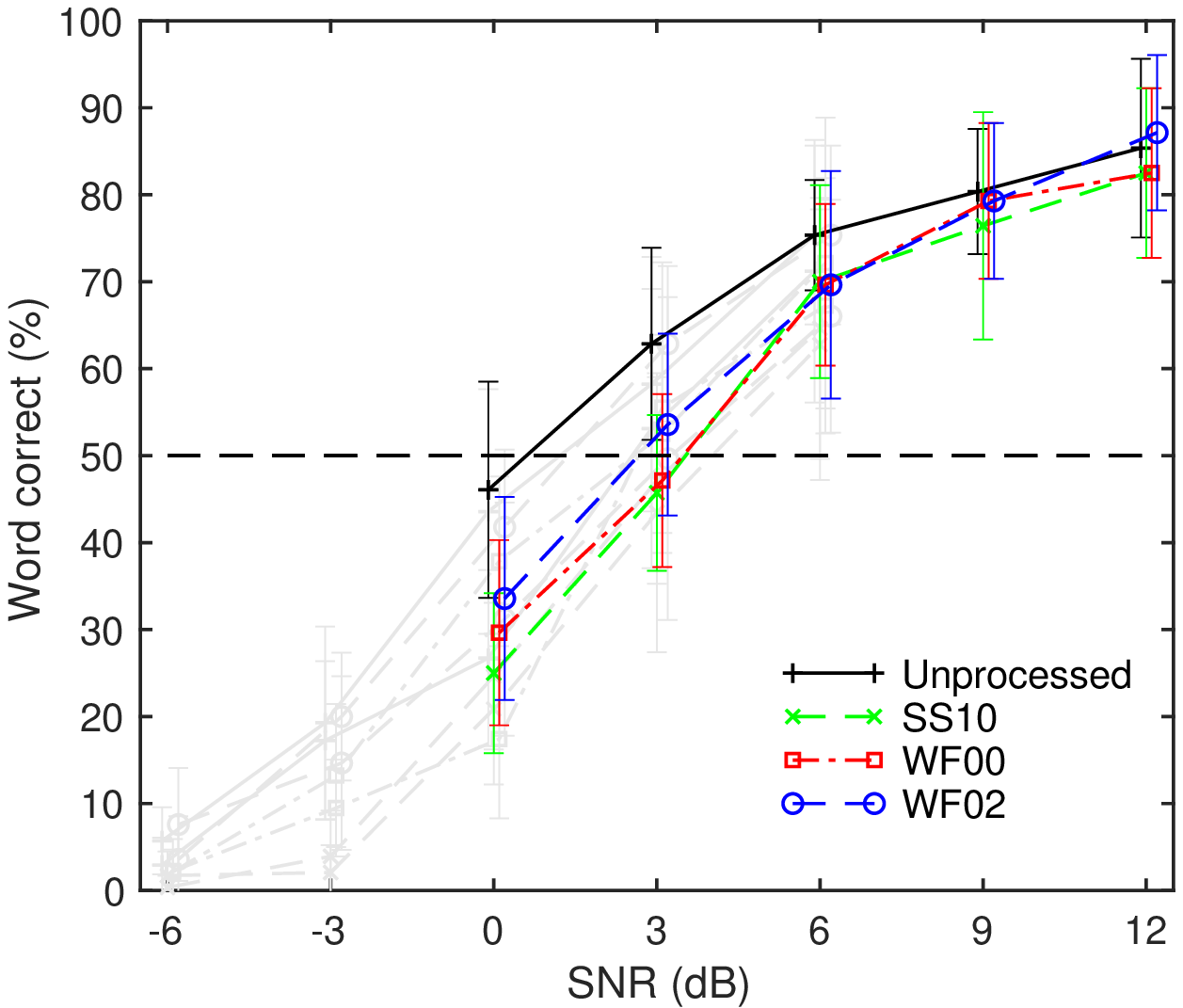}
          \caption{Laboratory experiment (low SPL, 43 dB $L_{\rm Aeq}$)\cite{Iwaki2019effect}. Mean and SD of word correct  (\%)   \\}
          \label{fig:ExpLabLowSPL_WCcurve}
        \end{center}
      \end{minipage}

    \end{tabular}
      \end{center}
      \vspace{-25pt}
\end{figure*}


\vspace{-6pt}
\section{Results}
\label{sec:ExpResult}
\vspace{-5pt}

The participants' responses in the remote experiments in section \ref{sec:ExpRemote} were compared with the results of the laboratory experiments \cite{yamamoto2020gedi,Iwaki2019effect} in section \ref{sec:ExpLab}.

\vspace{-8pt}
\subsection{Data cleansing}
\label{sec:DataCleansing}
\vspace{-5pt}
The remote experiment data consisted of lists of 4-mora words typed in by the participants.  We also collected scanned versions of the hand-written answer sheets to confirm that the answers had been entered correctly and to discourage the workers from cheating by giving irresponsible answers. All data were stored on the local website.

Data cleansing was performed in two steps.
The first step involves checking whether the 4-mora words the participants typed-in were entered correctly. This was done using a program, and any errors were corrected in accordance with the tendency of other participants' answers. The endeavor was not very time consuming. The second step is to compare the typed-in and hand-written words for whole answers. Any errors were also corrected, except when the answers were not interpretive. Few corrections were necessary. 
We also counted the number of hand-written corrections, which were probably made at the end of the session, since the participants were only allowed 4 seconds to write down the words during each listening period. As a result, we found that one participant had corrected their hand-written answers more than 80 times. We assumed that they did not understand the experimental instructions. We therefore excluded their results in both experiments from the analysis. Consequently, data for 29 participants were analyzed for each experiment.
The second step took 20 to 30 minutes per participant.
Although this was feasible for 30 participants, it would be difficult to accomplish given the large volume of data. However, we found that the following results were fairly unchanged, even without the second step.

\vspace{-8pt}
\subsection{Psychometric function of speech intelligibility}
\label{sec:SpIntel}
\vspace{-5pt}

The psychometric functions of word correct rates were calculated for each speech-enhancement condition as a function of the SNR. Figure \ref{fig:ExpRemote_WCcurve} shows the results of the remote experiments with 29 participants. The error bar represents the standard deviation (SD) across the participants. Figure \ref{fig:ExpLab_WCcurve} shows the results of the moderate SPL laboratory experiments with 14 participants. The faint gray lines behind the colored lines show the results of the other experiments. The lines were lower and the SDs were higher in Fig. \ref{fig:ExpRemote_WCcurve} than in Fig. \ref{fig:ExpLab_WCcurve}. 

Figure \ref{fig:ExpLabLowSPL_WCcurve} shows the results of the low SPL laboratory experiments with 14 participants. The lines between 0\,dB and 6\,dB were roughly overlapping, at least within the range of the SD, with those in Fig. \ref{fig:ExpLab_WCcurve}. As a result, even though the SPL difference was 20 dB, the effect of the SPL on speech intelligibility was not very large. It can be assumed that the listening levels in the remote experiments did not exert a significant effect on the participants' performance.

\begin{figure}[t] 
\vspace{-5pt}
\begin{center} 
    \centering
    \includegraphics[width = 1\columnwidth]
    {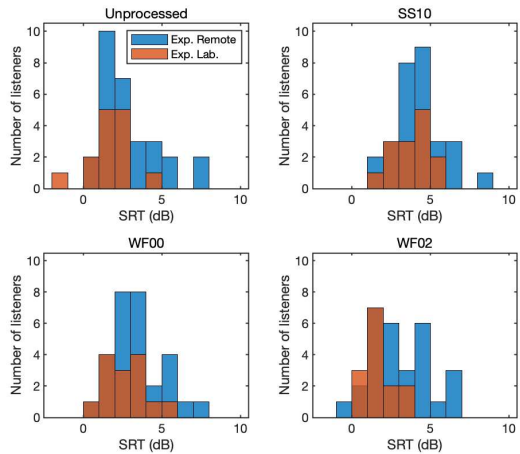}
    \vspace{-24pt}
      \caption{Histogram of SRT in dB for four speech-enhancement conditions. Blue: Remote experiment, Red: Moderate SPL laboratory experiment.
      }
      \label{fig:SRT_histogram}
\vspace{-28pt}
\end{center} 
\end{figure}

\begin{figure}[t] 
\vspace{-10pt}
\begin{center} 
    \centering
    \includegraphics[width = 0.8\columnwidth]
    {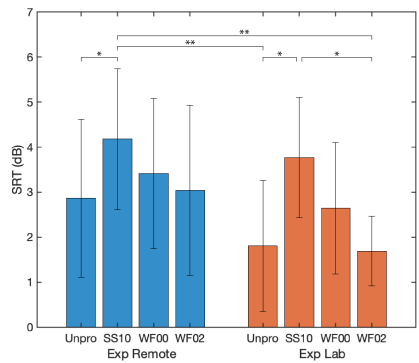}
        \vspace{-3mm}
      \caption{SRT in dB. Blue: Remote experiment, Red: Moderate SPL laboratory experiment.*: $p<0.05$, **: $p<0.01$.}
      \label{fig:SRT_Lab_Remote}
\vspace{-28pt}
\end{center} 
\end{figure}

\vspace{-8pt}
\subsection{Speech reception threshold}
\label{sec:SRT}
\vspace{-5pt}
Cumulative Gaussian psychometric functions were estimated from the data of the individual participants and the speech-enhancement conditions using a fitting procedure\cite{wichmann2001psychometric}. The speech reception threshold (SRT) is the SNR value where the psychometric function reaches a 50 \% word correct rate. 

Figure \ref{fig:SRT_histogram} shows histograms of the SRT values obtained in the remote experiments in Fig. \ref{fig:ExpRemote_WCcurve} (blue) and in the laboratory experiments in Fig. \ref{fig:ExpLab_WCcurve} (red) for each speech-enhancement condition. The peaks of the histograms were roughly the same, but the SRT distributions were extended to more than 5\,dB in the remote experiments. This is consistent with the higher SDs in Fig. \ref{fig:ExpRemote_WCcurve}. The larger variability may correspond to the diversity of the participants, whereas the participants in the laboratory experiments were restricted to young university students.

Figure \ref{fig:SRT_Lab_Remote} shows the mean and SD values of the SRTs dependent on the speech-enhancement conditions for the two experiments. Two-way analysis of variance (ANOVA) showed that there were significant main effects in the speech-enhancement conditions and the two experiments, but the interaction was not significant.
Multiple comparison analysis showed that there were significant differences ($p < 0.05 $) between ``Unprocessed'' and ``${\rm SS^{(1.0)}}$,'' but not between ``${\rm SS^{(1.0)}}$'' and  ${\rm WF_{PSM}^{(0.2)}}$ in the remote experiments. In contrast, there were significant differences for both cases in the laboratory experiments. This is the main difference between the two experiments. There are two significantly different conditions ($p < 0.01 $) across the two experiments, although the meaning of this is not easily interpreted. The important issue is that there were no significant differences between the other combinations.

The variations in the SRTs across the speech-enhancement conditions were similar between the two experiments. When developing and verifying a new objective speech intelligibility model, as in the case of GEDI \cite{yamamoto2020gedi}, ``Unprocessed'' was used as the reference condition to fix the parameter values, and the other speech-enhancement conditions were used to evaluate prediction performance. In this context, the results of the remote experiments could be usable, as well as the laboratory experiments.

\begin{figure}[t] 
\vspace{-5pt}
\begin{center} 
    \centering
    \includegraphics[width = 0.9\columnwidth]
    {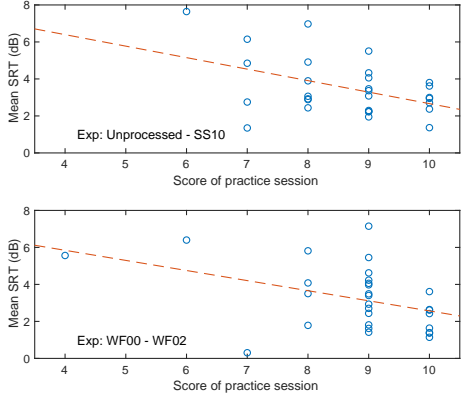} 
        \vspace{-10pt}
          \caption{Relationship between the practice session scores and the mean SRT (dB) with data (circles) and regression lines (dashed lines). Upper panel: First remote experiment; Lower panel: Second remote experiment}
          \label{fig:PredictSRT}
\vspace{-25pt}
\end{center} 
\end{figure}

\vspace{-8pt}
\subsection{Prediction of SRT}
\label{sec:DataScreening}
\vspace{-5pt}

To survey the factors influencing individual participants' SRT values, we performed stepwise regression analysis using generalized linear models. The target variable was the mean SRT of two speech-enhancement conditions, i.e., ``Unprocessed'' and ``${\rm SS^{(1.0)}}$'' in the first remote experiment and ``${\rm SS^{(1.0)}}$'' and  ${\rm WF_{PSM}^{(0.2)}}$ in the second remote experiment. We collected nine explanatory variables pertaining to participants' characteristics from the data registered on the crowdsourcing site and the experimental procedure: (1) age, (2) gender, (3) listening device (headphones or earphones), (4) reliability estimated from consistency of ID registration, (5) number of corrections of hand-written words, (6) number of inconsistencies between hand-written and typed-in words, (7) number of corrections due to a different mora count, (8) word correct score in the practice session (where 10 is a perfect score), and (9) duration of experiments.

The stepwise procedure, with a simple linear regression model, yielded simple equations as a function of the practice session scores. The equations for the first and second remote experiments were: 
\vspace{-4pt}
\begin{eqnarray}
  {\rm SRT (dB)}  & = &  8.88 - 0.63\times {\rm score}  \ \ (p = 0.015), \\
  {\rm SRT (dB)}  & = &  8.03 - 0.55\times {\rm score}  \ \  (p = 0.023).  
      \label{eq:SRTpred}
      \vspace{-10pt}
\end{eqnarray}
The linear models were significantly different from the constant models, and the coefficients were very similar.
The other factors were ruled out. The result implies that the practice session score is the only factor influencing the SRT values. 

Figure \ref{fig:PredictSRT} shows the SRT values and the regression lines. There was a clear tendency of negative correlation. The prediction errors were about 1.4\,dB and 1.6 \,dB, which are not very small but are comparable with the SDs shown in Fig. \ref{fig:SRT_Lab_Remote}.
\color{blue}

\color{black}

\begin{figure}[t] 
\vspace{-5pt}
\begin{center} 
    \centering
    \includegraphics[width = 0.9\columnwidth]
    {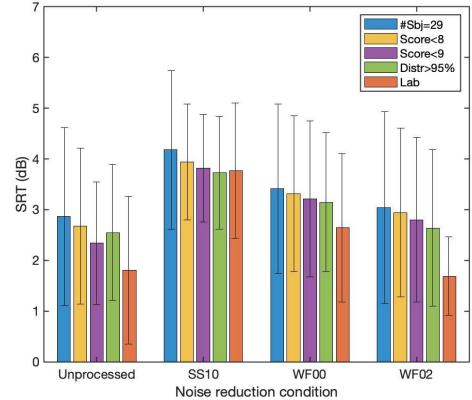} 
    \vspace{-10pt}
          \caption{Effect of data screening on the SRTs. Remote and laboratory experiments shown in Fig. \ref{fig:SRT_Lab_Remote} (blue and red);  Data elimination by the practice session score less than 8 (yellow) and less than 9 (purple), and by 
          cumulative Gaussian distribution greater than 95\% (green).  }
    \label{fig:SRT_DataElim}
\vspace{-25pt}
\end{center} 
\end{figure}

\vspace{-8pt}
\subsection{Data screening}
\label{sec:DataScreening}
\vspace{-5pt}

The practice session score is a priori information derived before performing the main tests. This is particularly important to judge whether the main tests are worth executing to obtain useful information. The score could be useful for data screening. If the practice session score is low, it may be that the participants did not fully understand the experimental procedure or they had difficulty filling in the words during the 4-second intervals of silence. There may be other reasons. 

We evaluated the effect of data reduction using the practice session scores. Firstly, the data of those participants who scored less than 8 were eliminated. Consequently, participant numbers were reduced to 24 and 26 in the first and second remote experiments, respectively. The result is shown in the yellow bars of Fig. \ref{fig:SRT_DataElim}. The mean and SD values of the SRTs reduced slightly compared to those in the original data with 29 participants (blue bars). Secondly, the data of those participants who scored less than 9 were eliminated. This reduced participant numbers to 17 and 22 in the first and second experiments, respectively. Again, the mean and SD of the SRTs dropped further but did not reach the level of the laboratory experiments (red bars). 

It is also possible to reduce the data after inspecting the response distribution. This seems to be common practice in sound quality assessment \cite{ribeiro2011crowdmos}. Initially, a Gaussian function was fitted to the SRT values shown in Fig. \ref{fig:SRT_histogram}, since a t-distribution with a degree of freedom of 28 is sufficiently close to the Gaussian. The samples above 95\% of the cumulative Gaussian distribution were then eliminated. The results are shown by the green bars in Fig. \ref{fig:SRT_DataElim}. The mean and SD values of the SRTs reduced slightly but, again, did not reach the level of the laboratory experiments (red). 
 
Consequently, it was difficult to select the remote data to be close to the laboratory data, probably because the two populations were different. However, it is worth noting that the a priori information about the practice session score works as well as the posteriori information about the distribution of the results.

\vspace{-10pt}
\section{Summary}
\vspace{-5pt}

In  this  study,  we  compared  speech  intelligibility results obtained in remote and laboratory experiments. Although the mean and SD of the SRT of the remote experiment were higher than those of the laboratory experiments, the variation in the SRTs across the speech-enhancement conditions was very similar between them. This implies that results of remote experiments may be usable to develop objective intelligibility measures, in addition to those of laboratory experiments. We also found that the a priori information about the practice session scores was useful for data screening to reduce the variance of the SRT.

\vspace{3pt}
\hspace{-12pt}
{\bf Acknowledgements }
\hspace{6pt}
This research was supported by JSPS KAKENHI Nos. JP16H01734 and JP18K10708.
 
\newpage
\bibliographystyle{IEEEtran}
  \newpage


%

\end{document}